\documentclass[twocolumn]{aastex631}
\usepackage{physics}
\usepackage{dsfont}
\shorttitle{Embedded disk size threaded by inclined field}
\shortauthors{Lee et al.}

\begin{document}


\title{Protoplanetary disk size under non-ideal magnetohydrodynamics: \\
A general formalism with inclined magnetic field}

\correspondingauthor{Yueh-Ning Lee}
\email{ynlee@ntnu.edu.tw}

\author[0000-0003-3497-2329]{Yueh-Ning Lee}
\affiliation{Department of Earth Sciences, National Taiwan Normal University, Taipei 11677, Taiwan}
\affiliation{Center of Astronomy and Gravitation, National Taiwan Normal University, Taipei 11677, Taiwan}
\affiliation{Physics Division, National Center for Theoretical Sciences, Taipei 10617, Taiwan}

\author[0009-0003-6115-1419]{Barshan Ray}
\affiliation{Department of Physical Sciences, Indian Institute of Science Education and Research Kolkata, Mohanpur 741246, India}
\affiliation{Department of Earth Sciences, National Taiwan Normal University, Taipei 11677, Taiwan}

\author[0000-0002-4577-8292]{Pierre Marchand}
\affiliation{Institut de Recherche en Astrophysique et Plan\'etologie, Universit\'e de Toulouse, UT3-PS, CNRS, CNES, 9 av. du Colonel Roche,
31028 Toulouse Cedex 4, France}

\author[0000-0002-0472-7202]{Patrick Hennebelle}
\affiliation{Universit\'{e} Paris Diderot, AIM, Sorbonne Paris Cit\'{e}, CEA, CNRS, F-91191 Gif-sur-Yvette, France}
\affiliation{IRFU, CEA, Universit\'{e} Paris-Saclay, F-91191 Gif-sur-Yvette, France}
\affiliation{LERMA (UMR CNRS 8112), Ecole Normale Sup\'{e}rieure, 75231 Paris Cedex, France}



\begin{abstract}
Many mechanisms have been proposed to alleviate the magnetic catastrophe, which prevents the Keplerian disk from forming inside a collapsing magnetized core. Such propositions include inclined field and non-ideal magnetohydrodynamics effects, and have been supported with numerical experiments. Models have been formulated for typical disk sizes when a field threads the rotating disk, parallel to the rotation axis, while observations at the core scales do not seem to show evident correlation between the directions of angular momentum and the magnetic field. In the present study, we propose a new model that considers both vertical and horizontal fields and discuss their effects on the protoplanetary disk size.
\end{abstract}



\section{Introduction}\label{sec:intro}
Due to advances in interferometric observations, the increasing number of embedded early phase disks now allows some interpretation of the physical process during the class 0/I young stellar object evolution phases. One major conclusion is that disks seem to be small with size not exceeding  50 AU at this stage. However, there have yet been strong conclusive results on whether the orientation of the magnetic field is correlated with the disk rotation axis \citep[see reviews e.g.][and references therein]{Maury2022r,Tsukamoto2023r}. The disk size at this stage is regulated by its magnetic interaction with the accreting envelope, and some scaling relations has been suggested assuming aligned field and rotation \citep{Hennebelle2016}. The ambipolar diffusion has been concluded to be the major effect regulating the disk size at longer timescales \citep{Lee2021b, Zhao2020r}. Numerical simulations suggested no correlation at the core scale \citep{Kuznetsova2020}. It is therefore important to understand how the inclination angle between the angular momentum and global magnetic field affects the formation of the disk. In this current work, we propose a model that describes the disk size as function of mass, magnetic field strength, and the misalignment angle, as a result of ambipolar diffusion. 

\section{Formalism}

\subsection{Properties of the envelope and the disk}\label{sec:prop}
The prescriptions follow exactly the same as that in \citet{Lee2021b}. 
The density in the collapsing envelope is proportional to the singular isothermal sphere \citep[SIS][]{Shu1977} 
\begin{eqnarray}\label{eq:rho_e}
\rho_{\rm e}(r) = {\delta_\rho C_{\rm s}^2 \over 2 \pi G r^2 },
\end{eqnarray}
where $r$ is the radial distance, $C_{\rm s} = 200~ {\rm m/s}$ is the thermal sound speed, $G$ is the gravitational constant, and $\delta_\rho$ is a numerical factor of the order of unity.
The density right inside the envelope-disk boundary is amplified by the accretion shock, such that
\begin{eqnarray}\label{eq:rho}
\rho(r) = {\delta_\rho C_{\rm s}^2 \over 2 \pi G r^2 } \mathcal{M}^2
= {\delta_\rho C_{\rm s}^2 \over 2 \pi G r^2 } \left(u_r(r)\over C_{\rm s}\right)^2,
\end{eqnarray}
with $\mathcal{M}$ being the Mach number. 
The Keplerian velocity is given by
\begin{eqnarray}\label{eq:v_kep}
v_{\rm kep} \simeq \sqrt{GM \over r} = \sqrt{G(M_\ast + M_{\rm d}) \over r},
\end{eqnarray}
where $M$ is the total mass of the system, with $M_\ast$ and $M_{\rm d}$ representing the mass of the star and that of the disk. 
The rotational velocity $u_\phi = \delta_\phi v_{\rm kep}$ and infall velocity $u_r = \delta_r v_{\rm kep}$ both scale with the Keplerian rotation velocity.
Vertical hydrostatic equilibrium results in the disk scale height
\begin{eqnarray}
h(r) \simeq C_{\rm s} \sqrt{ r^3 \over G M}. 
\end{eqnarray}
In a disk where the Keplerian rotation velocity is significantly supersonic, it follows naturally that $h \ll r$.
The numerical factors of order unity $\delta_\rho \gtrsim 1$, and $\delta_\phi, \delta_r \lesssim 1$.
It can be demonstrated that the results depend only sublinearly on these factors \citep{Lee2021b} and thus they will be ignored in the following discussions for conciseness.
In the present work, we discuss a general scenario where the magnetic field can be oriented in any direction. 
Assigning external field $B_0$ and field inclination angle $i$ with respect to the disk rotation axis, we can express the characteristic value of the three components as
\begin{eqnarray}
B_z &=& B_0 \cos{i},\label{eq:Bz} \\
B_r &=&B_0  {2 \over \pi}  \sin{i} , \label{eq:Br}\\
B_\phi &=& B_0 \chi, \label{eq:Bphi}
\end{eqnarray}
where $\chi$ is a dimensionless coefficient for the strength of the azimuthal component.
The factor ${2/\pi}$ in the radial component results from equally distributing the horizontal flux through the lateral cross-section of the disk into a cylindrical area, i.e., flux $B_0 2rh$ over area $\pi r h$. 
As a results of symmetry, the circular integral of $B_\phi$ along the disk edge is zero. The strength of the azimuthal field component is only a result of interplay between the magnetic flux and the disk rotation. 

\subsection{Nonideal MHD equations}
For treating the disk properties, we present below the non-ideal magnetohydrodynamic (MHD) equations. 
The equation of momentum conservation writes
\begin{eqnarray}\label{eq:momentum}
{\partial (\rho \vb*{u}) \over \partial t} = -\nabla \cdot\left[ \rho \vb*{u}\vb*{u} + \!\left( \!P \!+\! {B^2 \over 2} \right) \!\mathds{I} \!-  \vb*{B}\vb*{B} \right] -\rho \nabla \Phi,
\end{eqnarray}
where 
$\vb*{u}$ is the velocity vector, $P$ is the pressure, $\vb*{B}$ is the magnetic vector, and $\Phi$ is the gravitational potential.
The induction equation writes
\begin{eqnarray}\label{eq:B}
{\partial \vb*{B} \over \partial t} &=& \nabla \times (\vb*{u} \times \vb*{B} ) - \nabla \times \left\{ 
\eta_{\rm A} {\vb*{B} \over \| \vb*{B} \|} \times \left(\nabla \times \vb*{B}  \right) \times {\vb*{B} \over \| \vb*{B} \|}  \right.   \nonumber
\\
&+& \left. \eta_{\rm H} \left(\nabla \times\vb*{B}\right) \times {\vb*{B} \over \|\vb*{B}\|}
+ \eta_{\rm O} \nabla \times \vb*{B} \right\} \\
&=& \nabla \times \left[(\vb*{u}+\vb*{u}_{\rm A}+\vb*{u}_{\rm H}) \times \vb*{B} -\eta_{\rm O} \nabla \times \vb*{B}\right] \nonumber\\
&=& \nabla \times \left(\vb*{v} \times \vb*{B} - \eta_{\rm O} \nabla \times \vb*{B} \right),\nonumber
\end{eqnarray}
where $\eta_{\rm A}$, $\eta_{\rm H}$, and $\eta_{\rm O}$ are the resistivities of the ambipolar diffusion, Hall effect, and Ohmic dissipation, respectively.
The Hall effect is more relevant for transient disk behaviors \citet{Zhao2020r} and the Ohmic dissipation operates at higher densities. The present work focuses on the effects of the ambipolar diffusion. 


\subsection{Governing equations of the disk}

\citet{Lee2021b} discussed the strong-field and weak-field cases, where the diffusion of field lines is predominantly due to the magnetic tension and magnetic pressure gradient, respectively. This does not necessarily mean that the magnetic field is strong or not in absolute values, but actually indicates its strength with respect to the fluid inertia and the extent of field line winding due to induction. In reality, a growing disk quickly enters the weak field regime, where the azimuthal field generated by differential rotation is the dominant component, although the field strength could actually increase in time. 
Here we rename them magnetic-dominated and inertia-dominated regimes, which are more physically intuitive. 
The field lines tend to straighten themselves due to the magnetic tension in the magnetic-dominated regime. 
On the other hand, field lines tend to passively follow the rotating flow and have more winded configuration in the inertia-dominated regime. 

The angular momentum is maintained by the balance between the accretion from the envelope and the braking due to the magnetic tension, such that the azimuthal component of Equation (\ref{eq:momentum}) simplifies to
\begin{equation} \label{eq:AM}
0 \simeq -{u_r \over r} {\partial r u_\phi \over \partial r} + {\partial \left(B_z B_\phi \right)\over \rho\partial z}+ {\partial \left( r B_r B_\phi\right)  \over \rho r \partial r}. 
\end{equation}
The azimuthal magnetic field results from the balance between the induction from the vertical and radial components and the magnetic diffusion that redistribute the field lines. From Equation (\ref{eq:B}),
\begin{eqnarray}
0 &\simeq& {\partial u_{\phi} B_z \over \partial z} + {\partial u_{\phi} B_r \over \partial r} \label{eq:Bphi_t}\\
&+ &{\partial \over \partial z}\left[ {\eta_{\rm A} \over B^2} B_\phi \left( B_r {\partial B_r \over \partial z} + B_\phi {\partial B_\phi \over \partial z} \right) \right] +{\partial \over \partial z} \left( {\eta_{\rm A} B_z^2 \over B^2} {\partial B_\phi \over \partial z} \right).\nonumber
\end{eqnarray}
For conciseness, only the non-ideal terms with second derivative in $z-$direction are kept since they are the dominating terms in the disk geometry. 
The two diffusion terms correspond to the vertical diffusion due to pressure gradient and the azimuthal diffusion due to magnetic tension that tends to straighten the field lines. The non-ideal MHD diffusion in the radial direction is subdominant in the disk geometry \citep{Lee2021b} and is neglected in the following discussions.

\section{Results}

\subsection{Disk threaded by horizontal field: inertia-dominated regime}\label{sec:WB}
\citet{Lee2021b} and \citet{Hennebelle2016} only discussed disk threaded by an external vertical field. Here we start by discussing the case where the external field is purely horizontal. This implies that $B_z \ll B_\phi,B_r$ since the vertical velocity is negligible and no vertical component is generated through induction. 

In the regime where field lines are significantly winded, the azimuthal magnetic field is lost through diffusion due to vertical pressure gradient. This corresponds to the weak field case in \citet{Lee2021b}. Solving 
\begin{eqnarray}
{\rho u_\phi u_r \over r} = {B_\phi B_r \over r} \label{eq:AM_horizontal}
\end{eqnarray}
{from Equation (\ref{eq:AM}) and
\begin{eqnarray}\label{eq:Bphi_t_WB}
{u_{\phi} B_r \over  r} = {\eta_{\rm A} B_\phi  \over h^2}
\end{eqnarray}
from Equation (\ref{eq:Bphi_t}), we obtain 
\begin{eqnarray}\label{eq:rH_wB}
r_{\rm h} &=& \left[ (2\pi)^{-2}G^{3}C_{\rm s}^{-4} \eta_{\rm A}^{2} M^{5}\left(B_0{2\over\pi}\right)^{-4} \right]^{1\over11}\\
&=&  53.6 ~ {\rm AU}  
\left[{\eta_{\rm A} \over 10^{19} {\rm cm}^2~{\rm s}^{-1}}\right]^{2\over11} \left[{M \over 0.1 M_\odot}\right]^{5\over11}\left[ {B_0 \over 0.1 {\rm G}}\right]^{-{4\over11}}. \nonumber
\end{eqnarray}
When simplifying from Equation (\ref{eq:Bphi_t}) to Equation (\ref{eq:Bphi_t_WB}), it is presumed that $B^2 = B_\phi^2 + B_r^2$. Since we are already estimating properties at the disk edge without describing the exact profile, a factor $1/2$ that would result from the integration along $z$ is neglected for simplicity.
The field line straightening due to magnetic tension in the $r-\phi$ plane has a much larger timescale and is therefore not relevant for the equilibrium. 
In the scenario where this tension term dominates, a different solution can be found (see Appendix \ref{ap:SB}).

\subsection{Horizontal vs. vertical field}
Recall that the stationary disk size threaded by a vertical field is \citep{Lee2021b}
\begin{eqnarray}\label{eq:rV}
r_{\rm v} &=& \left[ (2\pi)^{-2}G^{1}C_{\rm s}^{0} \eta_{\rm A}^{2} M^{3}B_0^{-4} \right]^{1\over9}\\
&=&  19.2 ~ {\rm AU}  
\left[{\eta_{\rm D} \over 10^{19} {\rm cm}^2~{\rm s}^{-1}}\right]^{2\over9} \left[{M \over 0.1 M_\odot}\right]^{1\over3}\left[ {B_0 \over 0.1 {\rm G}}\right]^{-{4\over9}}. \nonumber
\end{eqnarray}
Combining Equations (\ref{eq:rH_wB}) and (\ref{eq:rV}), one derives the ratio
\begin{eqnarray}\label{eq:MB}
{r_{\rm h} \over r_{\rm v}} &=& \left[ (2\pi)^{4}G^{16}C_{\rm s}^{-36} \eta_{\rm A}^{-4} M^{12} B_0^{8} \right]^{1\over99} \left({2\over\pi}\right)^{-{4\over 11}}\\
&=&  2.80 
\left[{\eta_{\rm A} \over 10^{19} {\rm cm}^2~{\rm s}^{-1}}\right]^{-{4\over99}} \left[{M \over 0.1 M_\odot}\right]^{4\over33}\left[ {B_0 \over 0.1 {\rm G}}\right]^{8\over99}. \nonumber
\nonumber
\end{eqnarray}
This ratio is only weakly dependent on the disk parameters.
In the stellar mass regime, the disk size is always larger when threaded by a horizontal field. 

\subsection{Disk threaded by inclined field}
Without loss of generality, we replace the derivatives with characteristic values in Equations (\ref{eq:AM}) and (\ref{eq:Bphi_t}) to obtain the simplified system, which should still stay valid within correction by some numerical factor:
\begin{equation}\label{eq:AM_inclined}
{\rho u_\phi u_r \over r} \simeq {B_\phi B_z \over h} + {B_\phi B_r \over r}.
\end{equation}
\begin{equation}\label{eq:ind_inclined}
 {u_{\phi} B_z \over h} + {u_{\phi} B_r \over r} 
 \simeq {\eta_{\rm A} \left(B_\phi^2+B_r^2\right) B_\phi \over B^2 h^2} + {\eta_{\rm A} B_z^2 B_\phi \over B^2 h^2} 
= {\eta_{\rm A}  B_\phi \over h^2} ,
\end{equation}
Combining the two equations to eliminate $B_\phi$, the equilibrium radius will satisfy
\begin{equation}\label{eq:ri_general}
{B_z \sqrt{r}\over \sqrt{\eta_{\rm A} \rho u_r}} + {B_r h\over \sqrt{\eta_{\rm A} \rho u_r r}} = 1.
\end{equation}
This equation is general to any prescription of density and velocity profiles.
Inserting the disk properties from Sect. \ref{sec:prop} leads to
\begin{eqnarray}\label{eq:r_inclined}
\left({r \over r_{\rm v}}\right)^{9\over 4} \cos{i} + \left({r \over r_{\rm h}}\right)^{11\over 4} \sin{i}  = 1.
\end{eqnarray}
It is straightforward to show that the radius reduces to $r_{\rm v}$ or $r_{\rm h}$ when $i=0$ or $\pi/2$.

\begin{figure}
\includegraphics[trim=0 0 0 0,clip, width=0.5\textwidth]{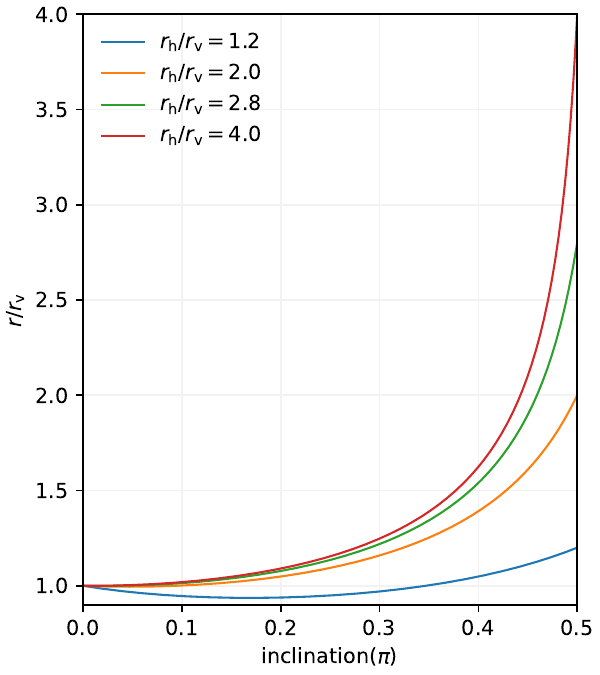}
\caption{The disk radius, divided by $r_{\rm v}$ plotted against the inclination angle $i$. Several values of $r_{\rm h}/r_{\rm v}$ are shown, while 2.8 is characteristic of the common disk parameters.\label{fig:r_inclined}}
\end{figure}

The results of disk radius as function of $i$ are shown in Figure \ref{fig:r_inclined} for several values of $r_{\rm h}/r_{\rm v}$. There exists a minimal disk size, $r_{\rm min}$, that is smaller than $r_{\rm v}$ (see derivation in Appendix \ref{ap:rmin}). This is particularly evident when $r_{\rm h}$ is close to $r_{\rm v}$. However, such minor effect might be difficult to detect through observations. 
The effect of horizontal field component only becomes significant at large inclinations. This might be able to explain that only some, while not many, large disks exist when the field inclination is indeed uniformly distributed. 

Combining Equations (\ref{eq:AM_inclined}) and (\ref{eq:ind_inclined}) allows to express $\chi$ in terms of the disk parameters.
The strength of the induced azimuthal field is
\begin{eqnarray}
\chi &=&\left[ r^{-5}  (2\pi)^{-2}G^{1}C_{\rm s}^{4} \eta_{\rm A}^{-2} M^{3} B_0^{-4} \right]^{1\over4}\\
&=& \left({r\over r_{\rm v}}\right)^{-{5\over4}} \left({r_{\rm v} C_{\rm s}\over \eta_{\rm A}}\right)  =
 \left({r\over r_{\rm v}}\right)^{-{5\over4}} \left[{r_{\rm v} \over 33.3 {\rm AU}}\right] .\nonumber
\end{eqnarray}
There is a critical value $33$ AU, below which $B_z > B_\phi$ for $i=0$. We recover what was derived in Table 2 of \citet{Lee2021b}. This is the separation between the magnetic-dominated and inertia-dominated regimes. 
In the perpendicular configuration $i=\pi/2$, lower mass or more strongly magnetized disks can be dominated by the radial field component. These conditions are less difficultly achieved in common disk environments. 
This is not unexpected given that the magnetic flux from the external field naturally gives rise to both radial and azimuthal components. 

The values of $\chi$ are plotted against the inclination $i$ for several values of $r_{\rm h}/r_{\rm v}$  in Figure \ref{fig:chi_inclined}. The black lines represent the strength of the vertical (solid) and radial (dashed) field components. The azimuthal field strength, presented with colored curves, is not only function of $r_{\rm h}/r_{\rm v}$, but also has to be rescaled according to the actual value of $r_{\rm v}$. Theses curves move upwards as $r_{\rm v}$ increases, and this is why large disks are always dominated by the toroidal field. 
\begin{figure}
\includegraphics[trim=0 0 0 0,clip, width=0.5\textwidth]{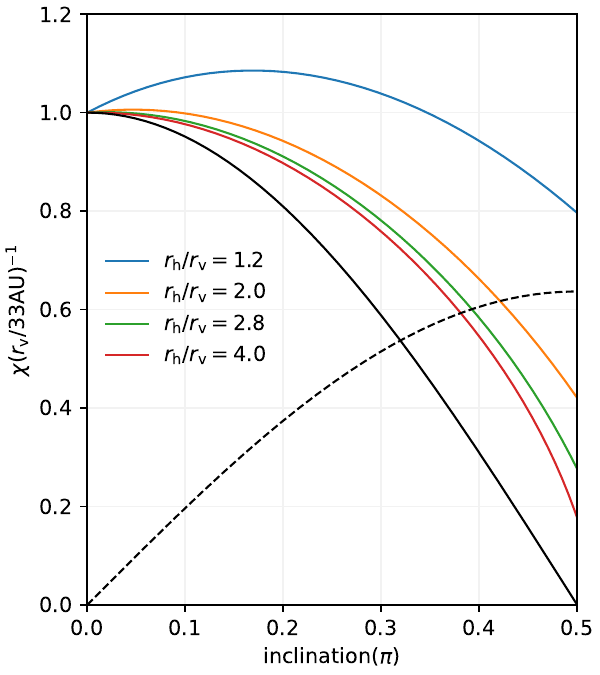}
\caption{Relative strength of the magnetic field components. Colored lines show $(r/r_{\rm v})^{-5/4}$, which are representative of the azimuthal field strength. Solid black line shows $\cos(i)$ ($z$-component) and dashed black line shows $\sin(i)2/\pi$ ($r$-component). The azimuthal field is dominant in most of the cases. \label{fig:chi_inclined}}
\end{figure}

We compared some simulation results from the literature and the difference with our model prediction is always within 30\%. For the aligned configuration, \citet{Tu2023} presented three cases with varied ambipolar diffusion strength, and we took values from their profiles of $B_z$ and $\rho$ at $M=0.25~M_\odot$ with their analytical formula for $\eta_{\rm A}$. \citet{Xu2021} showed profiles of $B_z$ and $\rho$ for four snapshots of varying accreted mass. Their ambipolar diffusion coefficient is not displayed thus we used the canonical value  $\eta_{\rm A}=10^{19} {\rm cm}^2~{\rm s}^{-1}$ as reference.  \citet{Masson2016} considered two initial values of magnetization. Only total $B$ is shown and we took the saturation value to approximate $B_z$ and also used the canonical value for $\eta_{\rm A}$.  The disk in the weakly magnetized case shows spirals much larger than the model size, while the more regular central part still fits quite well with our model prediction. On the other hand, their two cases misaligned by $40^\circ$, where the derived $r/r_{\rm v} \approx 1.1$, also agree within 20\% of the model prediction. \citet{Hennebelle2020a} presented a series of runs with varied initial magnetization and mass. The cases with $30^\circ$ misalignment are well described by the model with aligned field. While the value of $B_r$ is not presented, using the canonical value $B = 0.1$ G also reproduces very well their two cases with $90^\circ$ alignment at different levels of initial magnetization. 

\section{Discussion}

\subsection{Disk size population}
Our model describes the disk size as function of the inclination angle between the angular momentum and the external magnetic field. Observations of disk size distribution of a population can thus be used to test whether this angle $i$ is randomly distributed or follow some certain distribution $P(i)$. 
The numerical integration requires some special care since there exists a minimum radius, $r_{\rm min}$, at inclination $i_{\rm min}$.
By changing the variables, it is possible to perform the integration in $i$:
\begin{eqnarray}
N(<r) &=& \int\limits_{r_{\rm min}}^r{\mathrm{d}N \over \mathrm{d}r} \mathrm{d}r
= \int\limits_{r_{\rm min}}^r {\mathrm{d}N \over \mathrm{d}i} {\mathrm{d}i \over \mathrm{d}r} \mathrm{d}r \\
&=&
 \int\limits_{r_{\rm min}}^rP(i){\mathrm{d}i \over \mathrm{d}r} \mathrm{d}r =
\int\limits_{i_1(\min(r,r_{\rm v}))}^{i_2(r)} P( i) \mathrm{d}i,   \nonumber 
\end{eqnarray}
where $i_1(r)<i_{\rm min}$ and $i_2(r)>i_{\rm min}$ are the two values corresponding to the same radius $r$. 
\footnote{We derived such expressions to stay mathematically coherent in the discussions, while $r_{\rm min}$ is usually very close to $r_{\rm v}$ in the physically relevant regime (see Fig. \ref{fig:r_inclined}). This small difference is easily overwhelmed by uncertainties due to our simplification in some geometrical constants, and should not have strong physical implications.}
In Figure \ref{fig:r_pop}, we plot the number fraction of disks with size larger than $r$, which writes
\begin{eqnarray}
{\rm CDF}(>r) = 1-{N(<r) \over N(<r_{\rm h}) },
\end{eqnarray}
while assuming uniform distribution of inclinations, that is, $P(i) \propto \sin(i)$, or a more aligned distribution, $P(i) \propto 1$. 
As already pointed out in Figure \ref{fig:r_inclined}, the disk size only approaches $r_{\rm h}$ when the field is significantly inclined. Therefore, the size distribution is dominated by small ones even with a uniform inclination distribution, and worsens with a more aligned distribution. This effect is even more pronounced for higher mass and more magnetized disks (represented by large $r_{\rm h}/r_{\rm v}$). This is probably the reason why large disks are rare from observations \citep{Maury2022r}.

\begin{figure}
\includegraphics[trim=0 0 0 0,clip, width=0.5\textwidth]{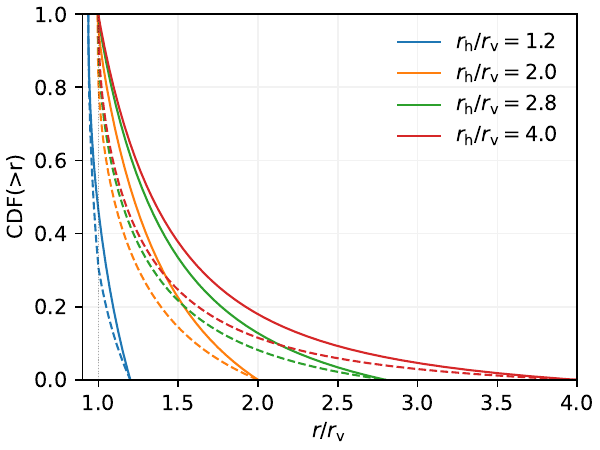}
\caption{Disk size distribution for varied values of $r_{\rm h}/r_{\rm v}$. Misalignment angle following uniform distribution, $P(i) \propto \sin(i)$, is plotted with solid lines, and a more aligned distribution,  $P(i) \propto 1$, is plotted with dashed lines. \label{fig:r_pop}}
\end{figure}

For the size distribution of the complete proto-stellar population, we integrate over the IMF and display the results in Figure \ref{fig:IMF}. We relate the disk parameters, including the field strength and $\eta_{\rm A}$, to the total mass in order to express everything with one single parameter (see Appendix \ref{ap:M}). With such parametrization, higher-mass stars have wider distribution of disk sizes. 
We integrated from $0.05$ to $300~M_\odot$, with a lognormal distribution around $0.2~M_\odot$ and powerlaw with slope $dN/d\log M \propto-1.35$ above $1~M_\odot$ \citep{Chabrier2005}. To examine the effect of the IMF variation, we also used a slope of $-0.75$  \citep{Lee2018a}. Three alignment distributions are presented: uniform ($P(i) \propto \sin (i)$, solid curves), moderately aligned ($P(i) \propto 1$, dashed curves), and perfectly aligned (dotted curves). 
The population is seriously dominated by the low-mass stars and thus no significant variation results from different IMF slopes. On the other hand, the disk size distribution seems to be a good proxy to test alignment between the rotation and external magnetic field. 
\begin{figure}
\includegraphics[trim=0 0 0 0,clip, width=.5\textwidth]{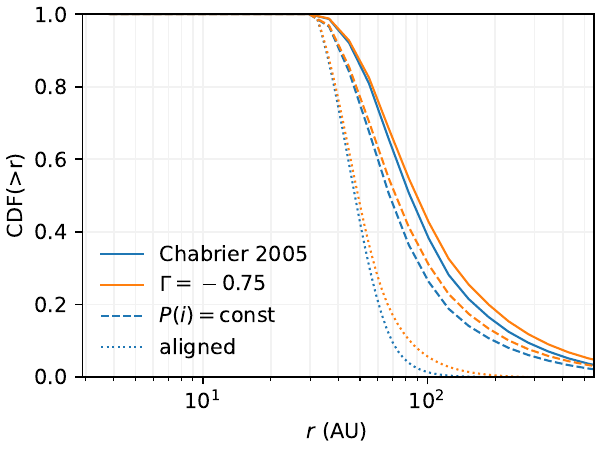}
\caption{Disk size distribution for a complete population following IMF distribution. Blue curves correspond to a population that follows a lognormal around $0.2~M_\odot$ and a powerlaw of slope $-1.35$ \citep{Chabrier2005}, and  orange curves have the IMF slope changed to $-0.75$. Misalignment angle following uniform distribution, $P(i) \propto \sin(i)$, is plotted with solid lines, and a more aligned distribution,  $P(i) \propto 1$, is plotted with dashed lines. \label{fig:IMF} }
\end{figure}

\citet{Lebreuilly2023} presented disk size distributions for series radiative non-ideal MHD simulations. Their results generally agree our model prediction for moderately aligned distribution. However, one case with lower initial global magnetization yields larger disks. This might come from the fact that in such conditions the disk magnetization does not follow our model simplification of field saturation regulated by non-ideal MHD (Appendix \ref{ap:M}). \citet{Bate2018} analyzed the disk size distribution in a clustered star formation stimulation, and they found surprisingly good agreement with observed class 0 disks without introducing magnetic fields. They identified dynamical encounters as an explanation for the compact disk sizes. The size distribution they discovered (their Fig. 28) falls in between our moderately aligned and perfectly aligned cases. These numerical works suggest that there could be multiple factor regulating the embedded disk size. Knowing complete disk parameters including mass, field strength, and alignment angle, rather than reducing to one single parameter, would definitely give more confidence in the prediction of disk sizes.  

\subsection{Role of the Hall effect}

In the early phases of disk formation, the Hall effect causes a bimodality in disk sizes, depending on the relative orientations of the angular momentum and the magnetic field \citep{Tsukamoto2015a}. 
However, as already demonstrated analytically in \citet{Lee2021b} and numerically in \citet{Zhao2020r}, this effect is transient and disappears after a few thousands years. 
Irrespective of the initial angular momentum, a disk with antiparallel angular momentum to the external field always forms at the end.
This is due to the conversion of poloidal field in toroidal field, that weakens or strengthens the magnetic braking in the infalling envelope. This is described with an additional term
\begin{equation}
    \frac{\partial B_\phi}{\partial t}_\mathrm{Hall} = -\frac{\partial}{\partial z}\left[\frac{\eta_\mathrm{H} B_z}{B}\frac{\partial B_r}{\partial z}\right]
\end{equation}
in the induction equation (\ref{eq:Bphi_t}).

When threaded by a horizontal field, the Hall drift of charged particles brings the field lines off the disk plane, which is described with the induction term
\begin{equation}
    \frac{\partial B_z}{\partial t}_\mathrm{Hall} = -\frac{\partial}{\partial r}\left[\frac{\eta_\mathrm{H} B_r}{B}\frac{\partial r B_\phi}{r\partial r}\right].
\end{equation}
This generated vertical field is parallel to the rotation in the envelope and anti-parallel inside the disk. 
The disk and the magnetic field lines will finally adjust to a configuration such that the rotation axis becomes anti-parallel to the local magnetic field. In the end, the quasi-stationary disk size mostly results from the effect of ion-neutral (ambipolar) friction. 


\section{Conclusions}
We proposed a general formalism for size of a disk threaded by inclined magnetic field, as a result of self-regulation of angular momentum and field strength by the non-ideal MHD effects, notably the ion-neutral friction. While we presented results with some prescriptions for density, velocity, and field strength, this formalism is directly applicable to any other profiles of these physical parameters. 
Our model allows to infer a disk size distribution from a distribution of the misalignment angle between rotation and magnetic field. We also reduced the disk size expression to function of the mass only. It is then possible to integrate over a stellar mass distribution function to obtain the size distribution of a complete protostellar population. Such model can be tested with size observations of a disk population.

\section*{Acknowledgments}
We would like to thank the anonymous referee for the thorough reading and constructive comments that helped to better shape the manuscript. YNL acknowledges funding from the National Science and Technology Council, Taiwan (NSTC 112-2636-M-003-001) and the grant for Yushan Young Scholar from the Ministry of Education, Taiwan.
PM received financial support from the European Research Council (ERC) under the European Union's Horizon 2020 research and innovation programme (ERC Starting Grant ``Chemtrip'', grant agreement No. 949278)
BR acknowledges financial support from the National Science and Technology Council, Taiwan (NSTC) under the International Internship Pilot Program (IIPP), and the Kishore Vaigyanik Protsahan Yojana (KVPY) program, India. 
PH acknowledges funding from the European Research Council synergy grant ECOGAL (Grant: 855130).

\appendix

\section{Magnetic-dominated regime with horizontal diffusion (strong field case)}\label{ap:SB}
In contrast to Sect. \ref{sec:WB}, we consider the possible scenario at the very beginning of disk formation, where the toroidal field is not well developed.
Under such conditions, the azimuthal component is removed by the field line straightening due to magnetic tension, instead of vertical diffusion.
In the absence of vertical field, the only magnetic tension comes from the product between the radial and azimuthal field components. The induction-generated azimuthal field is balanced by horizontal diffusion
\begin{eqnarray}
{\partial B_\phi \over \partial t} \simeq  {\partial u_{\phi} B_r \over \partial r} + {\partial \over \partial r}\left[ \eta_{\rm A}  {\partial \left( r B_\phi  \right) \over \partial r}\right].
\end{eqnarray}
By solving simultaneously Equation (\ref{eq:AM_horizontal}) and
\begin{eqnarray}\label{eq:Bphi_t_SB}
{u_{\phi} B_r \over  r} = {\eta_{\rm A} B_\phi  \over r^2},
\end{eqnarray}
we obtain the radius

\begin{eqnarray}\label{eq:rH_sB}
r_{\rm h} = \left[ (2\pi)^{-2}G^{1}C_{\rm s}^{0} \eta_{\rm A}^{2} M^{3}\left(B_0{2\over\pi}\right)^{-4} \right]^{1\over9} 
=  23.4 ~ {\rm AU}  
\left[{\eta_{\rm A} \over 10^{19} {\rm cm}^2~{\rm s}^{-1}}\right]^{2\over9} \left[{M \over 0.1 M_\odot}\right]^{1\over3}\left[ {B_0 \over 0.1 {\rm G}}\right]^{-{4\over9}}. 
\end{eqnarray}
This is almost identical to what was found by \citet{Hennebelle2016}, while the only difference comes from the rescaling due to Equation (\ref{eq:Br}). 
The right-hand side of Equation (\ref{eq:Bphi_t_SB}) comes from the radial diffusion due to gradient of $B_\phi$ and azimuthal diffusion due to the $B_\phi B_r$ tensor in Eq. (\ref{eq:B}), which are not shown in the simplified expression in Equation (\ref{eq:Bphi_t}). 
In disk geometry, the diffusion timescale, $r^2/\eta_{\rm A}$, is much larger than that in the inertia-dominated regime, $h^2/\eta_{\rm A}$. 
The magnetic tension diffusion is only relevant at the very early stage of prestellar core collapse, where a disk is not yet well developed. This solution is therefore not very relevant for a physical disk.


\section{Minimal disk size}\label{ap:rmin}
Taking the derivative of Equation (\ref{eq:r_inclined}) with respect to $i$,
\begin{eqnarray}
\left[{9\over 4}\left({r \over r_{\rm v}}\right)^{9\over 4} \cos{i} +{11\over 4} \left({r \over r_{\rm h}}\right)^{11\over 4} \sin{i}\right]  {\mathrm{d}r \over \mathrm{d}i}
-\left({r \over r_{\rm v}}\right)^{9\over 4} \sin{i} + \left({r \over r_{\rm h}}\right)^{11\over 4} \cos{i} = 0,
\end{eqnarray}
allows to find the minimum radius at which $dr/di=0$. 
The critical radius satisfies the condition
\begin{eqnarray}
\left({r_{\rm min} \over r_{\rm v}}\right)^{9\over 2}  + \left({r_{\rm min}  \over r_{\rm h}}\right)^{11\over2} = 1,
\end{eqnarray}
and is only slightly inferior to , $r_{\rm v}$.
The corresponding inclination
\begin{eqnarray}
i_{\rm min} = \arctan \left[  \left({r_{\rm min}  \over r_{\rm h}}\right)^{11\over4}  \left({r_{\rm min} \over r_{\rm v}}\right)^{-{9\over 4}} \right].
\end{eqnarray}


\section{Magnetic field strength and the stellar mass}\label{ap:M}
Due to the ambipolar diffusion, the magnetic field strength saturates at the center of a collapsing core.
In order to estimate this value at saturation for cores of various mass, we solve the radial equilibrium between advection and diffusion while assuming the radial component to be negligible
\begin{equation}
u_r B_z ={\eta_{\rm AD} \over B^2} \left( B_z^2{\partial B_z \over \partial r } + {B_\phi B_z \over r} {\partial r B_\phi \over \partial r }\right).
\end{equation}
For a shallow $B_\phi$ profile (shallower than $r^{-1}$), the second term on the right-hand side operates in synergy with the advection term. Therefore, the left-hand side can only be balanced by the gradient of $B_z$ component, whatever the dominant component is. When $B_\phi \gg B_z$, $B_\phi$ should obey a profile (probably close to $\propto r^{-1}$) such that the inward diffusion due to its gradient is dominated by the outward diffusion due to the gradient of $B_z$ and 
\begin{equation}
u_r B_z = \eta_{\rm AD} {B_z^2 \over B^2}{\mathrm{d}B_z \over \mathrm{d}r}
\end{equation}
always stays true.
Replacing the expressions with Equations (\ref{eq:rho_e}) and (\ref{eq:v_kep}) for a core that follows the description in Section \ref{sec:prop},
one obtains
\begin{eqnarray}
{\mathrm{d}B_z \over \mathrm{d}\rho} &=& {u_r B^2  \over \eta_{\rm AD}B_z} {\mathrm{d}r \over \mathrm{d}\rho} 
= {B^2  \sqrt{GM}\over B_z 2\rho  \eta_{\rm AD}}  \left({C_{\rm s}^2 \over 2\pi G\rho}\right)^{1\over4}.
\end{eqnarray}
The ambipolar diffusion coefficient is dominated by positively charged ions in the regime of our interest and thus
\begin{eqnarray}\label{eq:eta}
{4\pi \over c^2} \eta_A \approx \sigma_\perp^{-1} = {1+\left( \omega_{\rm M} \tau_{\rm Mn}\right)^2 \over \sigma_{\rm M}} 
 \approx \left\{ \begin{array}{rcl} {B^2 \over n_{\rm M} n c^2 m_{\rm H_2} a_{\rm MHe} \langle\sigma_{\rm coll}w\rangle_{\rm M}} ,&~ {\rm for} & n<n_{\rm crit} \\
{n  m_{\rm H_2} a_{\rm MHe} \langle\sigma_{\rm coll}w\rangle_{\rm M} \over n_{\rm M} e^2}, & ~{\rm for} & n>n_{\rm crit} \end{array} \right. ,
\end{eqnarray} 
where 
\begin{eqnarray}\label{eq:ncrit}
n_{\rm crit}  = {e \over c m_{\rm H_2} a_{\rm MHe} \langle\sigma_{\rm coll}w\rangle_{\rm M}} B =  k_{\rm crit} B
\end{eqnarray} 
is the density above which $\omega_{\rm M} \tau_{\rm Mn}<1$.
$\omega_{\rm M}$ is the cyclotron frequency of the ion, $\tau_{\rm Mn}$ is its characteristic time of collision with the neutral hydrogen, $c$ is the speed of light, $e$ is the elementary charge, and $m_{\rm H_2}$ is the mass of a hydrogen molecule. We adopt the same constants used in \citet{Marchand2016} for positive ions, $a_{\rm MHe} = 1.14$ \citep{Desch2001} is a constant factor and $\langle\sigma_{\rm coll}w\rangle_{\rm M} = 9.1 \times 10^{-10}$ is the rate constant for collision with H$_2$ molecules \citep{Pinto2008}. 
For the number density of the ions, we follow the fiducial case in \citet[][Fig.3]{Marchand2016} and parametrize as
\begin{eqnarray}\label{eq:nM}
n_{\rm M} = n_{\rm M0} \left({n \over n_{\rm ref}} \right)^\alpha,~{\rm with}~
\alpha  = 
 \left\{ \begin{array}{rcl} 0.3 ,&~ {\rm for} & n<n_{\rm ref} \\
0, & ~{\rm for} & n>n_{\rm ref} \end{array} \right. ,
\end{eqnarray} 
where $n_{\rm M0} = 10^{-2}$ cm$^{-3}$ and  $n_{\rm ref} = 10^{9}$ cm$^{-3}$. 
At high densities, the ion-electron recombination ($\propto n_\mathrm{H_2}^2$) becomes more efficient than their creation by cosmic rays ($\propto n_\mathrm{H_2}$), leading to an ion/electron fractional abundance scaling as $n_\mathrm{H_2}^{-1}$, and therefore a constant abundance \citep{Marchand2021}.
The system can be solved for three segments:
\begin{eqnarray}\label{eq:B_rho1}
B(n) =
 \left[{\left(GM\right)^{1\over 2} \over \left(\alpha+3/4\right)}\left( {C_{\rm s}^2 \over 2\pi G \mu m_{\rm p} }\right)^{1\over 4} 
4\pi m_{\rm H_2} a_{\rm MHe} \langle\sigma_{\rm coll}w\rangle_{\rm M}
{n_{\rm M0} n^{\alpha+3/4} \over n_{\rm ref}^\alpha } \right]^{1\over 2},
\end{eqnarray} 
for the two $\alpha$ values, separated at  $n_{\rm ref}$, up to density  $n_{\rm crit}$, and then
\begin{eqnarray}\label{eq:B_rho2}
B(n) = {n_{\rm crit} \over k_{\rm crit}}
\exp{\left[ {\left(GM\right)^{1\over 2} \over \left(5/4-\alpha\right)}\left( {C_{\rm s}^2 \over 2\pi G \mu m_{\rm p} }\right)^{1\over 4} 
{2\pi \over c^2}  {e^2 n_{\rm M0}\over m_{\rm H_2} a_{\rm MHe} \langle\sigma_{\rm coll}w\rangle_{\rm M}
n_{\rm ref}^\alpha }\right] \left(n_{\rm crit}^{\alpha-5/4}-n^{\alpha-5/4}  \right) }.
\end{eqnarray} 
The expression for density $n < n_{\rm crit}$ in Equation (\ref{eq:B_rho1}) is always true, irrelevant to the strength of $B_\phi$, since the dependence of $\eta_{\rm A}$ cancels out with the total field strength. However, Equation (\ref{eq:B_rho1}) is only true when negligible rotation is present and the vertical field component dominates throughout the whole domain. In the presence of significant toroidal field, vertical equilibrium has to be solved simultaneously. If $B_\phi$ saturates at some value, a very shallow profile decreasing toward high densities will be recovered. 
We plot the $B-\rho$ relations for several values of core mass in Figure \ref{fig:B_rho}. 
This $B-\rho$ relation is derived for the collapsing envelope, and is therefore also valid for the scenario with field lines perpendicular to the rotation axis. 

The third segment is almost flat.
By combining Equations (\ref{eq:ncrit}) and (\ref{eq:B_rho1}), we can obtain the saturated field strength value
\begin{eqnarray}\label{eq:B_sat}
B_{\rm sat} \approx
\left[ {\left(GM\right)^{1\over 2} \over \left(\alpha+3/4\right)}\left( {C_{\rm s}^2 \over 2\pi G \mu m_{\rm p} }\right)^{1\over 4} 
{4\pi e \over c} k_{\rm crit}^{\alpha-{1\over 4}}  {n_{\rm M0}\over n_{\rm ref}^\alpha }\right]^{1/({5\over 4}-\alpha)} = 5.5 \times 10^{-2} ~{\rm G} \left[{M \over 0.1 M_\odot}\right]^{2\over 5}. 
\end{eqnarray} 
The values are consistent with MHD simulations \citep{Masson2016, Commercon2022}, and the results of \citet[][Figure 1]{Hennebelle2016} is essentially recovered with this analytical formalism without need to perform numerical integration.

\begin{figure}
\centering
\includegraphics[trim=0 0 0 0,clip, width=.5\textwidth]{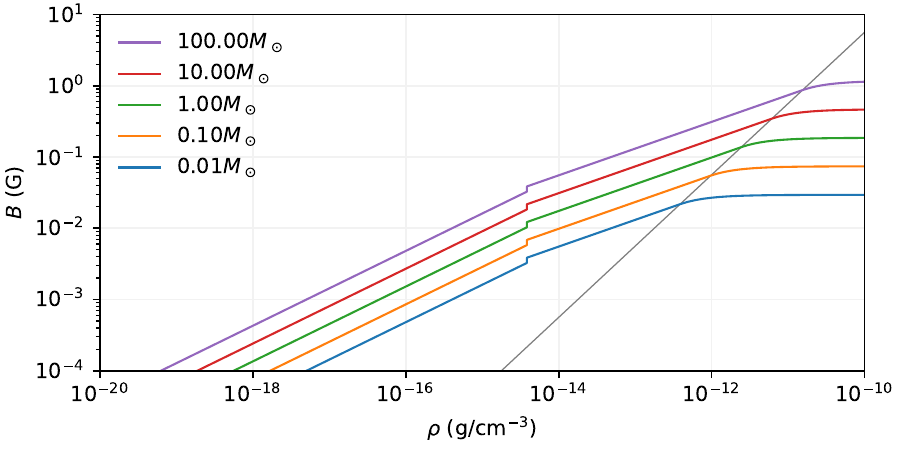}
\caption{Magnetic field strength-density relation for collapsing envelope around various central stellar mass, from Equations (\ref{eq:B_rho1}) and (\ref{eq:B_rho2}). The gray line shows the relation from Equation (\ref{eq:ncrit}) which seperates different regimes of $\eta_{\rm A}$ behaviors. The kink is due to the discontinuous expression in Equation (\ref{eq:nM}), which can be easily removed by applying some smoothing.\label{fig:B_rho}}
\end{figure}


While we leave $\eta_{\rm A}$, $B_0$, and $M$ and independent parameters in the main text, one can also take into account Equations (\ref{eq:eta}) and (\ref{eq:B_sat}), and apply Equation (\ref{eq:eta}) with $n>n_{\rm crit}$ when solving Equation (\ref{eq:ri_general}). This leads to \begin{eqnarray}
\left({r \over r_{\rm v}}\right)^{15\over 4} \cos{i} + \left({r \over r_{\rm h}}\right)^{17\over 4} \sin{i}  = 1,
\end{eqnarray}
where 
\begin{eqnarray}
r_{\rm v} = 26.8 ~{\rm AU} \left[{M \over 0.1 M_\odot}\right]^{17\over 75} 
~{\rm and}~ 
r_{\rm h} = 50.2 ~{\rm AU} \left[{M \over 0.1 M_\odot}\right]^{27\over 85}.
\end{eqnarray}
The density corresponding to this solution is much below the presumed magnetic field saturation density, and is thus not a self-consistent solution. Therefore, instead one should consider Equations (\ref{eq:eta}) and (\ref{eq:B_rho1}) when solving Equation (\ref{eq:ri_general}). 
By using $B \approx B_0\chi$ in Equation (\ref{eq:eta}) with $n<n_{\rm crit}$, this leads to 
\begin{eqnarray}
\left({r \over r_{\rm v}}\right)^{1-{9\over4}\alpha} \cos{i} + \left({r \over r_{\rm h}}\right)^{{3\over 2}-{9\over4}\alpha} \sin{i}  = 1,
\end{eqnarray}
where 
\begin{eqnarray}
r_{\rm v} =  \left\{ \begin{array}{l}  4.50 ~{\rm AU} \left[{M \over 0.1 M_\odot}\right]^{1\over 13} \\
 37.8 ~{\rm AU} \left[{M \over 0.1 M_\odot}\right]^{1\over 4} \end{array} \right.
~{\rm and}~ 
r_{\rm h} = \left\{ \begin{array}{l} 334 ~{\rm AU} \left[{M \over 0.1 M_\odot}\right]^{21\over 33} \\
199 ~{\rm AU} \left[{M \over 0.1 M_\odot}\right]^{1\over 2} \end{array} \right..
\end{eqnarray}
Whether the solution is valid depends on the corresponding density. The actual radius is the larger among the two values. 
This radius is subject to the upper limit imposed by a pure hydrodynamic treatment, which conserves all accreted angular momentum \citep{Lee2021b}:
\begin{eqnarray}
r_{\rm hydro} = 111 ~ {\rm AU} \left[{\beta \over 0.02}\right]^{1\over 2} \left[{M \over 0.1 M_\odot}\right],
\end{eqnarray}
where $\beta$ is the ratio between the rotational and gravitational energy in the core. This hydrodynamic radius could be over-estimated, since some braking should already happen within the envelope, particularly in the horizontal field case. Strictly speaking, there is an inconsistency because the density and the field strength experience an amplification across the shock at the disk edge. 
However, as shown by \citet[][Fig. 3]{Hennebelle2020a}, the amplification of $B_z$ is weaker than the amplification of the density. Using Equation (\ref{eq:B_rho1}) for the sublinear dependence of $B_z$ on $\rho$ turns out to be a reasonable estimate, even across the shock, due to its diffusive nature.

\bibliography{inclined_disk_final}{}
\bibliographystyle{aasjournal}



\end{document}